\documentclass[useAMS,usenatbib]{mn2e}

\newcommand{\msun}{\mbox{${\rm M}_{\odot}$} }

\newcommand{\bv}{Brunt-V\"{a}is\"{a}l\"{a} }
\usepackage{graphicx}
\usepackage{amsmath,amssymb}
\usepackage{natbib}
\def\mnras{MNRAS}
\def\apj{ApJ}
\def\aap{A\&A}
\def\aaps{A\&AS}

\def\apjs{ApJS}
\def\araa{ARA\&A}
\def\aapr{A\&AR}
\def\apss{Ap\&SS}%

\title[g-modes in supergiants: mass loss and overshooting]{Can mass loss and overshooting prevent the excitation of g-modes in blue supergiants?}
\author[M. Godart, A. Noels, M.-A. Dupret]{M. Godart$^{1}$\thanks{E-mail:
melanie.godart@ulg.ac.be} A. Noels$^1$ and M.-A. Dupret$^2$\\
$^1$ Inst. d'Astrophysique et de G\'eophysique -
Univ. de Li\`{e}ge, All\'ee du 6 Ao\^{u}t 17 - B 4000 Li\`{e}ge, Belgium\\
$^2$ Observatoire de Paris, LESIA, CNRS UMR 8109, 5 place J. Janssen, 92195 Meudon, France}

\begin{document}

%
\pagerange{\pageref{firstpage}--\pageref{lastpage}} \pubyear{2002}

\maketitle

\label{firstpage}

\begin{abstract}
Thanks to their past history on the main sequence phase, supergiant massive stars develop a convective shell around the helium core. This intermediate convective zone (ICZ) plays an essential role in governing which g-modes are excited. Indeed a strong radiative damping occurs in the high density radiative core but the ICZ acts as a barrier preventing the propagation of some g-modes into the core. These g-modes can thus be excited in supergiant stars by the $\kappa$-mechanism in the superficial layers due to the opacity bump of iron, at $\log T=5.2$. 
However massive stars are submitted to various complex phenomena such as rotation, magnetic fields, semiconvection, mass loss, overshooting. 
Each of these phenomena exerts a significant effect on the evolution and some of them could prevent the onset of the convective zone. 
We develop a numerical method which allows us to select the reflected, thus the potentially excited, modes only. We study different cases in order to show that mass loss and overshooting, in a large enough amount, reduce the extent of the ICZ and are unfavourable to the excitation of g-modes.
\end{abstract}


\section{Introduction}
\begin{figure}
\begin{center}
\includegraphics[width=65mm,angle=270]{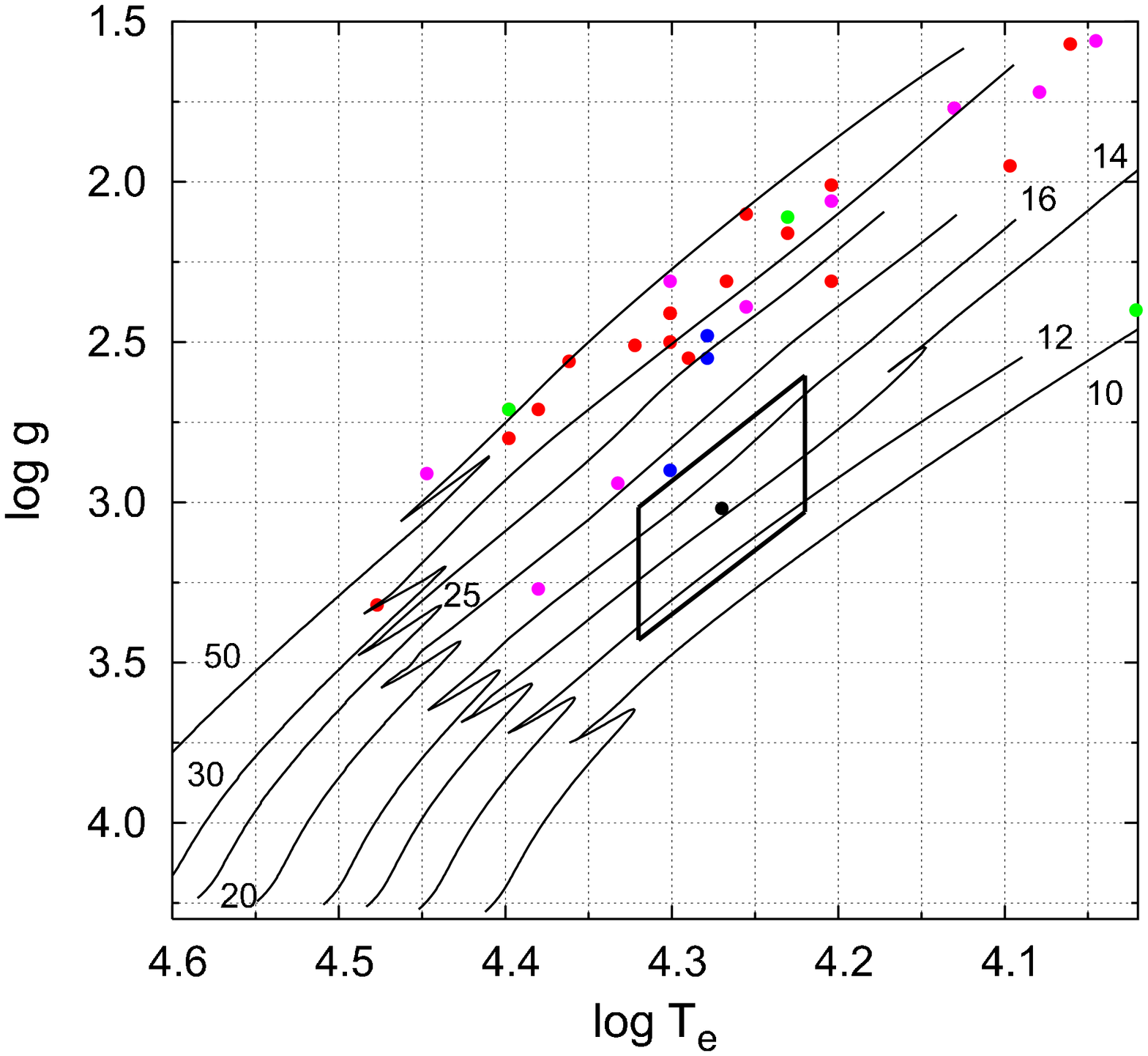}
 \caption{$\log g -\log T_{\sf eff}$ diagram in which the sample of periodically variable B type supergiants from \citet{Lefever2006} and some evolutionary tracks from $10$ to $50\msun$ computed with CLES are displayed. The MOST star is also shown with the error box derived from photometric studies \citep{Schmidt1996} by \citet{Saio2006}. \citet{Lefever2006} have divided their sample into 4 groups depending on the reliability of the derived stellar parameters (mainly $T_{\sf eff}$ and $\log g$). Red points are the results considered as very reliable, restricted reliability is shown in blue. Green points stand for the stars for which they had no means to derived accurate stellar parameters, the results are thus considered as unreliable and must be taken with caution. They used 12 comparison stars in which 9 of them appear to be periodically variable too (magenta points).}
 \label{fig:logg-logTe}
\end{center}
\end{figure}

Hipparcos data set has provided new samples of periodically variable B-type stars. Among them, about 30 periodically variable supergiant stars have been detected \citep{Waelkens1998,Aerts1999,Mathias2001}. \citet{Lefever2006} have recently reexamined this sample and they derived the atmospheric parameters through line profile fitting. From the location in the ($\log{g}$, $\log{T_{\sf eff}}$) diagram (Fig.~\ref{fig:logg-logTe}), they suggest that the variability is due to non-radial pulsations excited by the $\kappa$-mechanism. \citet{Saio2006} also reported the discovery of p and g-mode pulsations in a B supergiant star HD\,163899 (B2 Ib/II \citealt{Klare1977,Schmidt1996}) which has been observed by the MOST satellite. For this star 48 frequencies have been detected ($\lesssim 2.8$ c/d) with amplitudes of a few milli-magnitudes.

At first sight, the presence of g-mode pulsations in a supergiant star is quite unexpected. Indeed such stars present a rather condensed radiative  helium core with a very large \bv frequency which produces an important radiative damping. This radiative damping is so strong that no g-modes entering the core should be excited. However \citet{Saio2006}  have shown that the presence of excited g-modes is indeed possible thanks to an intermediate convective zone (ICZ) which prevents some of the modes from entering the radiative damping core. In that case, the $\kappa$-mechanism in the superficial layers can be sufficient to excite the modes. The presence of this ICZ is therefore crucial to have excited g-modes in supergiant stars.

However massive stars are submitted to various complex phenomena such as rotation, strong mass loss, magnetic fields, semi-convection and overshooting. Each of these phenomena exerts a significant effect on the evolution and some could prevent the formation of an ICZ. An enlightening and still widely used review on massive star evolution is to be found in \citet{Chiosi1986} while a more recent review has been made by \citet{Maeder1998}. 
We shall here limit our discussion to slow rotators, slow enough to be able to ignore the additional mixing induced by rotation. Magnetic fields will also be ignored. 

We shall here consider models on the main sequence (MS) or close to it and analyze the effect of mass loss and/or overshooting.
Mass loss induces a faster receding convective core during the MS phase \citep{Chiosi1986}. If a large enough mass loss rate is taken into account during the main sequence, no ICZ can be formed during the post-MS phase. Overshooting will be discussed according to two different aspects: (1) overshooting during MS can prevent the formation of an ICZ during the supergiant phase; (2) with a large amount of overshooting a star located in the supergiant region could still be on the MS, i.e. with a convective hydrogen-burning core. 
Although the location of these stars in the ($\log{g}$, $\log{T_{\sf eff}}$) diagram could also be attributed to stars undergoing central helium-burning at the blue part of the He loop, we shall concentrate in this paper on the H burning phases: either the main sequence or the supergiant phase.

\section{Models}

The models were computed with the Code li\'egeois d'Evolution Stellaire, CLES \citep{Scuflaire2007}. 
We used the OPAL opacities \citep{Iglesias1996}, completed with the \citet{Alexander1994} opacities at low temperature. We adopted the CEFF equation of state \citep{Christensen-Dalsgaard1992} and the old standard heavy-element mixture (Grevesse \& Noels 1993).
Boundary conditions at the surface are derived from a radiative gray model atmosphere computed in the Eddington approximation and with the OPAL equation of state. 
The adiabatic oscillation frequencies
are computed with LOSC \citep{Scuflaire2007a} and the excitation of oscillation modes is computed using the non-adiabatic code MAD \citep{Dupret2003}.

\section{Role of an ICZ in the excitation of g-modes in supergiant stars}
\label{sect:ICZ}

\subsection{Internal structure}
B supergiant stars are post-MS massive stars which are burning hydrogen in a shell surrounding the helium core. Some of them are already burning helium in the core. The He core is very dense and is contracting whereas the low density radiative envelope is expanding. The structure of the star in the vicinity of the hydrogen burning shell depends on the structure of the star during the MS. In particular, depending on this past history, an intermediate convective zone can develop above the core.

\begin{figure}
\begin{center}
\includegraphics[width=65mm,angle=270]{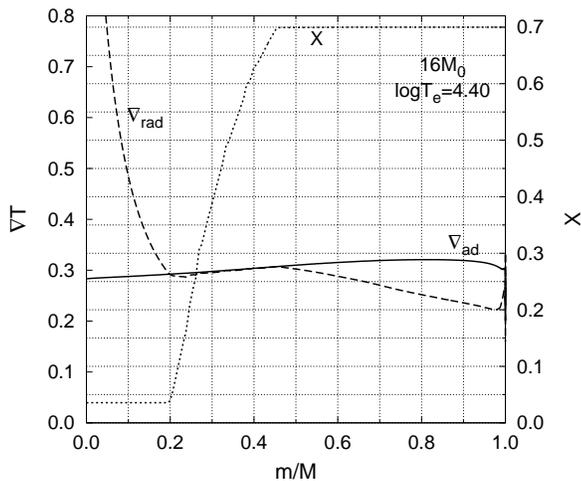}
 \caption{Radiative and adiabatic temperature gradients (dashed and solid line respectively) in a model of $16\msun$ on the MS phase. The dotted line stands for the hydrogen abundance. During the MS both the radiative and the adiabatic gradients decrease and the region with the neutrality of gradients is formed.}
 \label{fig:M16-gradX-article}
\end{center}
\end{figure}

The formation of the ICZ requires a region where the neutrality of temperature gradients is reached: $\nabla_{\sf rad} \cong\nabla_{\sf ad}$ during the MS phase.	
This region is formed by the decrease of the radiative and the adiabatic temperature gradient during the MS.
The transformation of H into He decreases the opacity and thus the radiative temperature gradient which, in less massive stars, leads to a receding convective core. However the larger the initial mass, the higher the central temperature and massive stars have therefore a large radiation pressure which increases during MS. This results in a decrease of the adiabatic temperature gradient and the convective core either recedes much more slowly or even increases in mass during the MS. When this happens, a semi-convective zone appears, and adopting the Schwarzschild criterion for convective stability, a partial mixing occurs in a region surrounding the convective core with a radiative temperature gradient very close to the adiabatic one (Fig.~\ref{fig:M16-gradX-article}). 
In the post-MS phase, when H-shell burning starts in this region, the radiative gradient becomes very rapidly larger than the adiabatic one since $L/m$ becomes very large ($\nabla_{\sf rad}\propto \kappa \,L/m$) and an intermediate convective zone appears in, or near, the H-burning shell.

However that MS 'neutral' region as well as the post-MS ICZ can be affected by various physical processes occurring  during MS.  If a large enough mass loss rate is taken into account during the MS phase, 
the central temperature increases less quickly. Hence, the adiabatic gradient no longer decreases
significantly. Mass loss also results in a convective core receding more quickly and does not leave behind it
a region where  $\nabla_{\sf rad}\cong\nabla_{\sf ad}$ \citep{Chiosi1986}. Therefore, 
with significant mass loss, no ICZ appears during the post-MS phase.
Overshooting during MS can also prevent the formation of an ICZ during the supergiant phase.
Finally, the criterion used to define the boundaries of the convective zones: 
Schwarzschild $\nabla_{\sf rad}=\nabla_{\sf ad}$ or Ledoux $\nabla_{\sf rad}=\nabla_{\sf ad}+(\beta/(4-3\beta))\nabla_{\sf \mu}=\nabla_{\sf L}$
also affects the appearance or not of an ICZ \citep{Lebreton}.

\subsection{Excitation of g-modes}
\begin{figure}
\begin{center}
\includegraphics[width=65mm,angle=270]{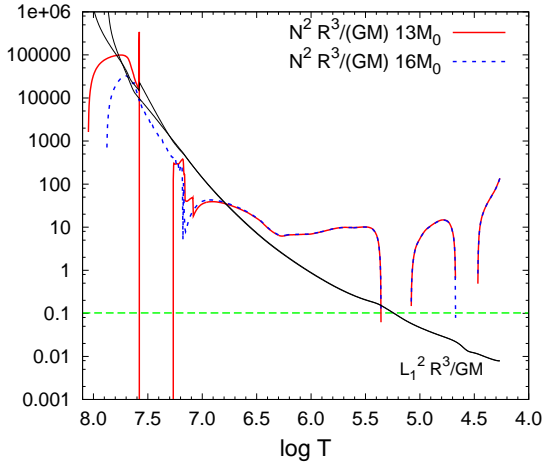}
 \caption{Propagation diagram for two supergiants model ($13\msun$, $\dot{M}=0$, $\log{T_{\sf eff}}=4.26$, $\log{L/L_{\odot}}=4.50$, solid line) and  ($16\msun$, $\dot{M}=10^{-7}\msun$/yr, $\log{T_{\sf eff}}=4.26$, $\log{L/L_{\odot}}=4.47$, dotted line). The squared dimensionless Lamb frequencies ($L_{\ell}^{2}\, R^3/(GM)$; similar for both models) are shown by solid lines for $l=1$. The squared dimensionless \bv frequencies ($N^2\, R^3/(GM)$) are shown by solid and dotted lines for the model with ICZ (without mass loss) and without ICZ (with mass loss) respectively. The horizontal dashed line corresponds to a typical SPB type g-mode.}
 \label{fig:propagation-diagram}
\end{center}
\end{figure}

The structure of a supergiant star is similar to the one of a red giant: it is characterized by a high density contrast between the small size core and the expanding envelope. This problem has been discussed for giant stars by \citet{Dziembowski1971,Dziembowski1977,Dziembowski2001}. The regions of mode trapping are shown in the propagation diagram (Fig.~\ref{fig:propagation-diagram}) in which the dimensionless squared \bv ($N^2$) and Lamb ($L_{\ell}^2$) frequencies are plotted as a function of $\log T$. We first consider a model without ICZ (dotted line).
The behaviour of $L_{\ell}=\sqrt{l(l+1)}\,c/r$ is qualitatively similar from star to star, being infinite at the center and decreasing monotically towards the surface. This is not the case for the \bv frequency:
\begin{eqnarray}
N^{2}&=&-g\bigg[\frac{d\ln\rho}{dr}-\frac{1}{\Gamma_{1}}\frac{dln P}{dr}\bigg]\\
&\simeq&\frac{g^{2}\rho}{P}\bigg[\nabla_{\sf ad}-\nabla+\nabla_{\mu}\bigg] \,\,\textrm{for a fully ionized gas}
\end{eqnarray}
which is affected by the presence of convective region where $N^{2}$ is close to zero and by the mean molecular weight gradient, $\nabla_{\mu}$ which leads to bumps in the \bv frequency. Moreover, the \bv frequency depends on $g^{2}\rho /P$, which in supergiant stars, due to the contrast between the contracting helium core and the expanding envelope, takes huge values within the radiative core. 
Hence the eigenmodes of any moderate frequency ($\omega \lesssim{100}$) show a g-mode behaviour in the core, where $\sigma\ll N,L_{\ell}$: all non-radial modes are mixed modes. Moreover, the eigenfunctions have a large wavenumber, $k$, in the core since:
\begin{eqnarray}
k^{2}&=&\frac{(N^{2}-\sigma^{2})(L_{\ell}^{2}-\sigma^{2})}{(\sigma{c})^{2}}\nonumber \\
&\simeq&\dfrac{N^{2}}{\sigma^{2}}\,\dfrac{l(l+1)}{r^{2}}\quad\rm{if}\,\,\sigma\ll N,L_{\ell}.
\end{eqnarray}
In this radiative g-mode cavity, where oscillations present short wavelengths, a strong radiative damping always occurs. 
Indeed, locally, the mechanical energy lost 
per unit length by the mode during each pulsation cycle due to radiative heat exchanges, is approximately 
given by:
\begin{eqnarray}
\quad
{\sf d} W_{\sf rad}/{\sf d} r &\cong& \frac{L}{\sigma\:{\sf d}\ln T/{\sf d}r}
\frac{{\delta T}}{T}
\frac{{\sf d}^2(\delta T/T)}{{\sf d}r^2}\nonumber\\ 
&=&
\ell (\ell+1)\:\frac{N^2}{\sigma^3}\:\frac{T^4}{\kappa\rho}\:\left|\frac{\delta T}{T}\right|^2\:\frac{16\pi a c}{3}\,.
\end{eqnarray}
This expression will be detailed more rigorously in Sect.~\ref{sect:quasiad}. It is obtained by substituting one of the dominating term in Eq.~\ref{diff} (last term) into Eq.~\ref{eq:sigmaI}. This radiative damping of non-radial modes is thus very large due to the huge values of $N^2$ in the core,
and it seems that there are no hopes to observe such modes in post-MS stars. 

\begin{figure}
\begin{center}
\includegraphics[width=58mm,angle=270]{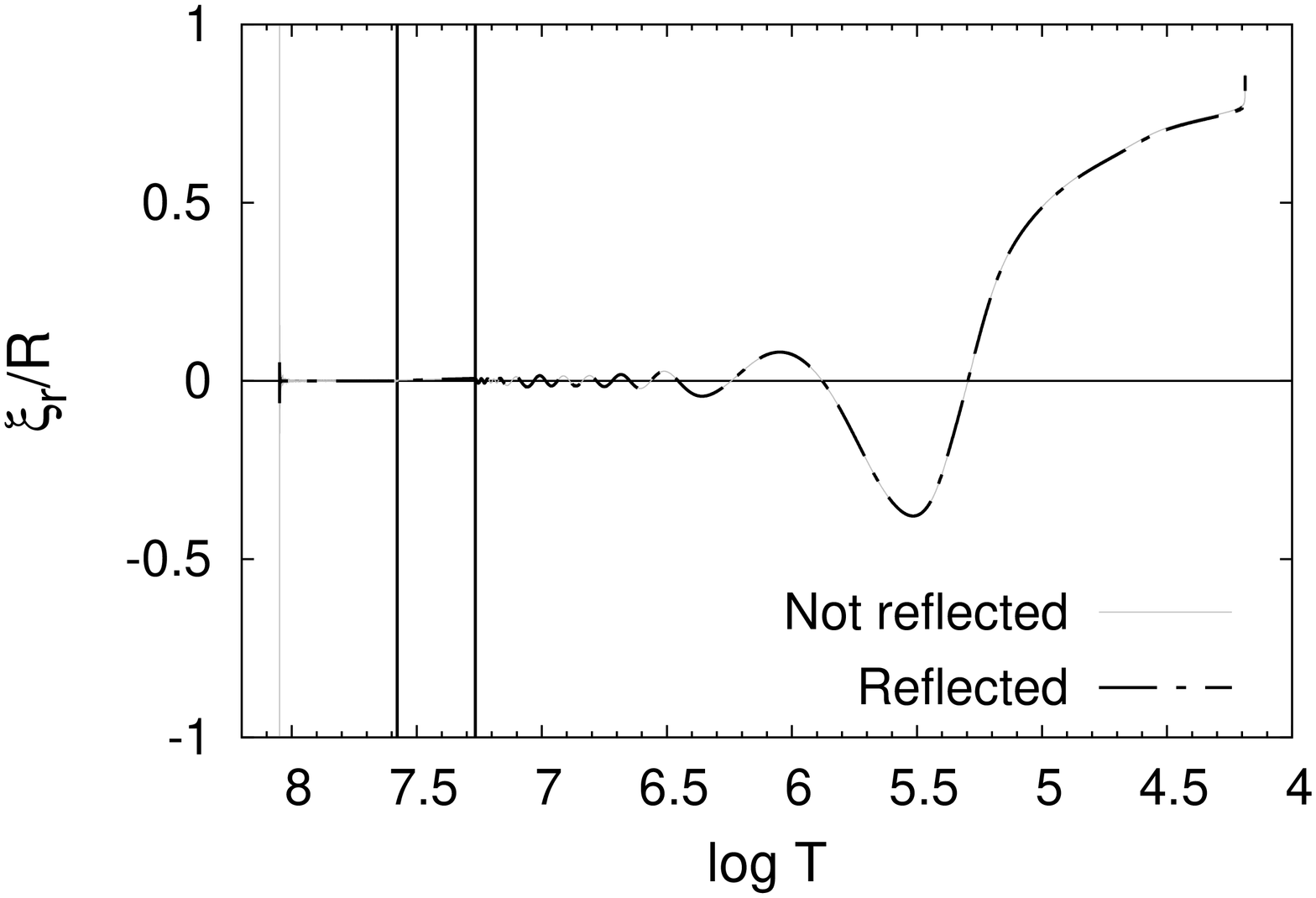}
\includegraphics[width=58mm,angle=270]{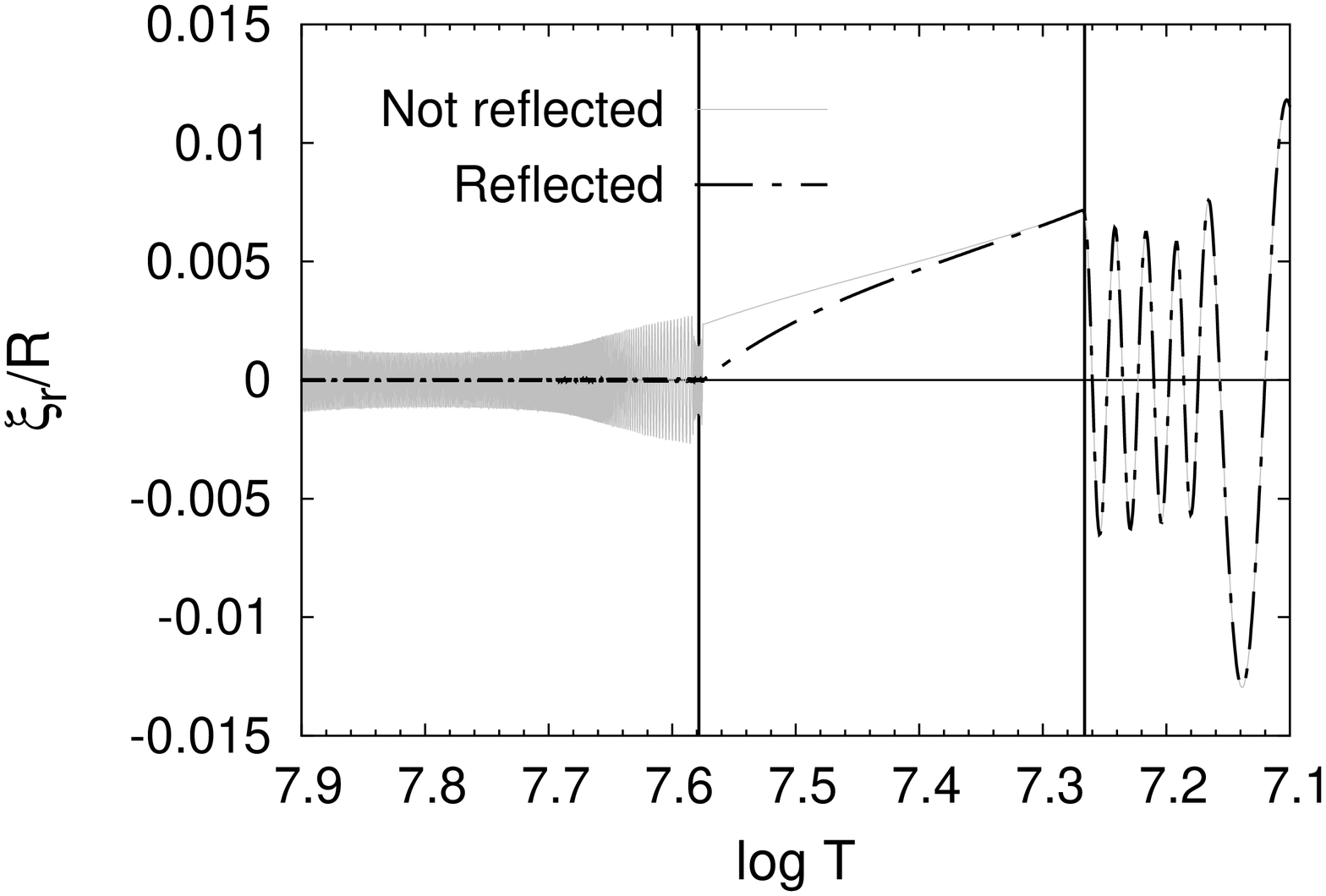}
 \caption{Radial displacement eigenfunction versus the temperature of the star. The ICZ is shown by vertical solid lines. Two modes are shown, the gray one crosses the ICZ and enters the radiative core in which it has large amplitude, whereas the other (black dashed line) is reflected on the ICZ. In that case, the amplitude in the core is small. The amplitudes in the envelope are roughly the same. The behaviour of the eigenfunctions in the center is an artefact coming from our use of the asymptotic approximation (Eq.~\ref{eq:xir}, where $N$ and $r$ come to zero). \textbf{Top} Representation of the whole star. \textbf{Bottom} Zoom on the central regions.}
 \label{fig:drzoom-M13-Mdot-0-0418-om0.29-ref-noref1-2}
\end{center}
\end{figure}

\begin{figure}
\begin{center}
\includegraphics[width=65mm,angle=270]{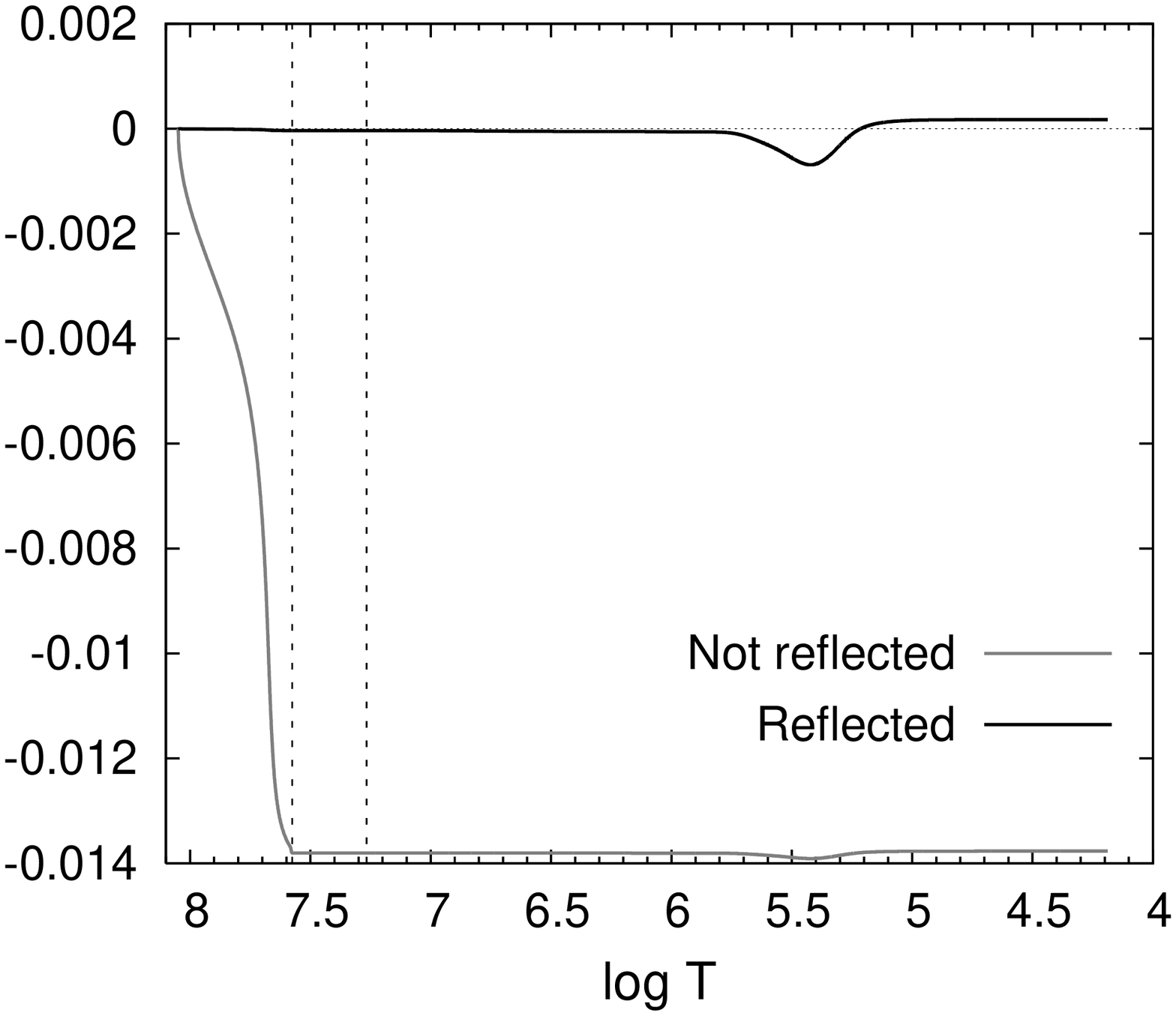}
 \caption{Work integral versus the temperature of the star. The work is positive at the surface if the mode is excited. The reflected mode (in black) is thus excited whereas the non-reflected one (in gray) suffers the strong damping in the radiative core.}
 \label{fig:workintegral}
\end{center}
\end{figure}

However, in the reasoning above we have assumed that the eigenfunction square modulus 
$\left|\delta T/T\right|^2$ is not negligible, which is not always the case due to the presence
of an ICZ (Fig.~\ref{fig:propagation-diagram}, solid line). In a convective zone $N^2 < 0$ and g-modes are evanescent. Therefore, the ICZ can act as a potential
barrier: some modes can cross it, others are reflected. This is illustrated in  Fig.~\ref{fig:drzoom-M13-Mdot-0-0418-om0.29-ref-noref1-2}.
For the mode that crosses the convective barrier (gray line), the amplitudes of short wavelength oscillations
are significant in the radiative core and strong radiative damping ensues (see the gray line on the work integral, shown on Fig.~\ref{fig:workintegral}): this mode 
cannot be observed. The other mode, whose frequency is very close, is reflected on the convective barrier, i.e. has small amplitudes in the radiative core
(black line). In this case the radiative damping remains small compared to the $\kappa$-mechanism occurring
in the iron opacity bump near the surface (see the increasing work at $\log T\simeq 5.2$, black line on Fig.~\ref{fig:workintegral}); 
this mode is unstable and could be observed. \textit{In the dense spectrum of non-radial
modes, some are reflected on the convective shell and can be excited and observed.} 
In order to select the reflected modes, i.e. those with small amplitudes in the radiative core, we impose a rigid boundary condition at the base of the ICZ (see Sect.~\ref{section:nummethode}). If the wavelength of the considered mode is a lot smaller than the width of the ICZ, this condition allows to achieve small amplitudes in the radiative core.
It is useful to note that,
if only reflected modes are considered, the mode propagation cavity exactly looks like a MS B star:
because of the reflection, the radiative core can be somehow forgotten and the bottom of the cavity is a convective region.
As a consequence, the frequencies of reflected modes behave like frequencies in a MS star. Limited to the reflected 
modes, the frequency pattern is sparse enough and asteroseismology of blue supergiants becomes possible.

\section{Supergiant stars: Numerical method}
\label{section:nummethode}

The main numerical problem when solving the non-adiabatic pulsation equations 
in supergiant stars is the large number of mixed modes with a huge number of 
nodes. Indeed, for a given period range, the number of nodes, $n$,  depends 
on the \bv frequency:
\begin{equation}
P_{\sf n,l}\simeq \frac{2\pi^{2}(n+\frac{l}{2}+\epsilon)}
{\sqrt{l(l+1)}\int_{0}^{r} \frac{N}{r}\,dr}\,.
\end{equation}
For a $\beta$ Cephei type mode with a period of the order of the fundamental 
radial mode, the number of nodes is of the order of 100, or even 1000 for SPB 
type g-modes.
The problem is numerical: it is difficult to solve the differential equations in such a 
cavity with so large a number of nodes. We shall however see that it is 
possible to avoid the computation of the full dense spectrum of non-radial 
modes by preselecting the modes that are reflected at the bottom of the ICZ
(and potentially excited).

First of all, in order to select the reflected modes only, we impose a rigid 
boundary condition at the bottom of the convective shell: $\xi_{r}=0$. This 
condition leads to small amplitudes of all eigenfunctions in the radiative 
core since $\vert P'\vert$ is much smaller than $\vert\delta{P}\vert$. As 
a consequence, $\vert\delta{P}\vert\simeq\rho g \vert\xi_{r}\vert$ stays also 
small. This condition allows us to solve the 'envelope problem' individually, 
without taking care of what is happening below. All the computations are 
performed with the standard fully non-adiabatic code. Artificially imposing a 
boundary condition inside the star will of course not provide the real 
frequencies of
the full star. But as the frequency spectrum of the full star is very dense, 
there always exists a mode
with a frequency close to the one corresponding to such a rigid reflection and 
the error introduced by this approach
is very small.

Once the non-adiabatic pulsation equations are solved in the envelope, we 
cannot completely disregard
the radiative core because reflection is only partial. In the radiative core, 
we have seen that the 
eigenfunctions have a huge number of nodes (up to $n\approx 1000$!). This 
makes very difficult to accurately solve
the differential problem in this region. But at the same time, two 
approximations are valid
with high accuracy in this region.

\subsection{Quasi-adiabatic treatment}
\label{sect:quasiad}

Firstly, the internal energy of these deep layers is very high, which justifies
the so-called quasi-adiabatic approximation. This approximation can be 
presented in different ways \citep{Dziembowski1977,Unno1989}, but leading always 
to the same final result.
We can present it also as follows. 
First, the adiabatic problem is solved to get the eigenfunctions $\xi_r$, 
$\delta P$,
\ldots \, Next, these adiabatic eigenfunctions are used to determine $\delta L$ 
from the perturbed diffusion equation:
\begin{equation} \label{diff}
\frac{\delta L}{L} \:=\: 4\,\frac{\xi_r}{r} \:+\:  
3\, \frac{\delta T}{T}
 \:-\: \frac{\delta\kappa}{\kappa} \:-\: \ell(\ell+1)\:\frac{\xi_h}{r} 
\:+\:\frac{{\rm d}\,\delta T/{\rm d} r}{\mathrm{d}T/\mathrm{d}r}\:.
\end{equation}
Finally, with all these ingredients the work integral can be computed
and divided by the inertia to get the damping rate of the modes, which gives 
in a pure radiative zone
without nuclear reactions:
\begin{equation}
\label{eq:sigmaI}
\eta\;=\;\frac{\int_{0}^{M}\frac{\delta T}{T}\left(\frac{\partial\delta L}
{\partial m}-
\frac{\ell(\ell+1)\,L}{4\pi\rho r^3}\frac{T'}{{\rm d}T/{\rm d}\ln r}\right)\,
{\rm d}m}
{2\:\sigma^2\int_{0}^{M}|\vec{\xi}|^2\,{\rm d}m}\,,
\end{equation}where $\sigma$ (resp. $\eta$) are the real (resp. imaginary) parts of the 
angular frequency (time-dependence: $\exp(i\sigma t - \eta t)$).
The quasi-adiabatic treatment is used for the integration in the radiative 
core ($r<r_0$) and the full non-adiabatic eigenfunctions are used for the 
envelope ($r\geq r_0$).  

\subsection{Asymptotic treatment}

Secondly, another important simplification is to use the asymptotic theory in 
the radiative core.
This theory applies perfectly there since the wavelength of the eigenfunctions 
is by far smaller than the scale heights of different equilibrium quantities. 
A full non-adiabatic asymptotic treatment
was derived by \citet{Dziembowski1977}. Here we use instead the standard adiabatic 
asymptotic theory,
which gives the following expressions for the radial ($\xi_r$) and transversal 
($\xi_h$) components 
of the displacement far from the edges of a cavity \citep{Unno1989}
\begin{eqnarray}
\label{eq:xir1}
\frac{\xi_{r}}{r}&=&\frac{K}{r^{2}\,\sqrt{c\,\rho\,\sigma}}\left[ 
\frac{L_{\ell}^{2}-\sigma^{2}}{N^{2}-\sigma^{2}}\right]^{1/4}\sin\left[ 
\int_{r_{0}}^{r}{k_{r}\,\rm{d}r}\right] \\
\label{eq:pprim1}
\frac{\xi_{h}}{r}&=&\frac{P'}{\sigma^2 r^2 \rho} \nonumber \\
&=&\frac{K}{r^{3}}\,
\sqrt{\frac{c}{\sigma^3\rho}}\,\left[\frac{N^{2}-\sigma^{2}}
{L_{\ell}^{2}-\sigma^{2}}\right]^{1/4}\cos\left[ \int_{r_{0}}^{r}
{k_{r}\,\rm{d}r}\right].
\end{eqnarray}
In a g-mode cavity ($\sigma^2 << L_\ell^2,\:N^2$), this gives:
\begin{eqnarray}
\label{eq:xir}
\frac{\xi_{r}}{r}&=&K\:\frac{[\ell(\ell+1)]^{1/4}}{\sqrt{\sigma\,r^5\,\rho\,N}}\;
\rm{sin}\left[ \int_{r_{0}}^{r}{k_{r}\,dr}\right]  \\
\label{eq:pprim}
\frac{\xi_{h}}{r}&=&\frac{P'}{\sigma^2 r^2 \rho}
\nonumber \\
&=& \frac{K}{[\ell(\ell+1)]^{1/4}}\:\sqrt{\frac{N}{\sigma^{3}\, r^5 \,\rho}}
\;\;\rm{cos}\left[ \int_{r_{0}}^{r}{k_{r}\,dr}\right]\,,
\end{eqnarray}
where the local radial wavenumber is given by $k_r=\sqrt{\ell(\ell+1)}\,N\,/\,
(\sigma\,r)$.
The $K$ constant is obtained by applying the continuity of $\delta P$ at the 
bottom of the ICZ ($r_0$),
which is equivalent to the continuity of $P'$ because of our rigid boundary 
condition. 
We note from these equations that $|\xi_h|/|\xi_r| \approx N/\sigma >> 1$ and
$|P'|/|{\rm d}P/{\rm d}r \,\xi_r| \approx N\sigma\, r/g \approx \sigma/L_\ell 
<< 1$.  
Hence $\delta P/P\:\simeq\:(d \ln P/dr)\: \xi_r$ (the same for $T$ and $\rho$) and  
$|\delta \rho/\rho|<<\ell(\ell+1)\:|\xi_h/r|$ (near incompressibility), so 
that 
$d\xi_r/dr\simeq \ell(\ell+1)\:\xi_h/r$.

Substituting Eqs.~\ref{eq:xir} and \ref{eq:pprim} in Eq.~\ref{diff} and 
keeping only the
dominating terms in the asymptotic limit gives thus:
\begin{eqnarray} \label{diff-as}
\frac{\delta L}{L} &\simeq& 
\frac{{\rm d}\,(\delta T/T)}{\mathrm{d}\ln T}
\:-\: \ell(\ell+1)\:\frac{\xi_h}{r}\nonumber \\
&\simeq&
\ell(\ell+1)\left(\frac{\nabla_{\sf ad}}{\nabla}-1\right)\,\frac{\xi_h}{r}\:.
\end{eqnarray}
We obtain then from the equation of energy conservation and neglecting the 
transversal component
of the divergence of the flux (since ${\rm d}\ln T/ {\rm d}r\,\,\rm{and}\,\,k_{r}\gg l(l+1)/r$):
\begin{eqnarray}
\label{deltas}
i\sigma T \delta s &\simeq&-\frac{d\delta L}{d m}\nonumber\\
&\simeq&K\;\frac{[\ell(\ell+1)]^{5/4}}{4 \pi}\,L\,\left(\frac{\nabla_{\sf ad}}
{\nabla}-1\right)\nonumber \\
& &\sqrt{\frac{N^3}{\rho^3 \sigma^{5} r^{11}}}\;\; \rm{sin}\left[ \int_{r_{0}}^{r}
{k_{r}\,dr}\right]\nonumber\\
&\simeq&\frac{\ell(\ell+1)}{4 \pi}\left(\frac{\nabla_{\sf ad}}
{\nabla}-1\right)
\frac{N^2\,L}{\sigma^2 \:r^3 \rho}\;\frac{\xi_r}{r}\:.
\end{eqnarray}

The contribution of the radiative core to the numerator of 
Eq.~(\ref{eq:sigmaI}) 
is thus simply given by:
\begin{eqnarray}
\label{numerator}
&&\int_{0}^{m_0}\frac{\delta T}{T}\frac{\partial\delta L}{\partial m}\:{\rm d} 
m \nonumber \\
&&\;\simeq\;\frac{\ell(\ell+1)}{\sigma^2}\:\int_0^{r_0} \frac{\rho g}{P}
\:\frac{\nabla_{\sf ad}-\nabla}{\nabla}\:\nabla_{\sf ad} N^2 
L\left(\frac{\xi_r}{r}\right)^2{\rm d} r\nonumber\\
&&\simeq\;\frac{K^2\,\left[\ell(\ell+1)\right]^{3/2}}{2\sigma^3}
\int_0^{r_0} 
\:\frac{\nabla_{\sf ad}-\nabla}{\nabla}\:
\frac{\nabla_{\sf ad}\,N\,g\,L}{P\:r^5}\;{\rm d} r
\end{eqnarray}
and for the denominator:
\begin{equation}
\label{denominator}
2\sigma^2\int_{0}^{m_0}|\vec{\xi}|^2\,{\rm d}m\;\simeq\;
4\pi\,K^2\,\int_0^{r_0} k_r\: {\rm d}r\:.
\end{equation}

These equations are perfectly compatible with those given in \citet{Van1998} and \citet{Dziembowski2001}.
Finally, it is important to emphasize that using Lagrangian or Eulerian 
formalisms can lead to different numerical results. In our first 
computations, we used
a Lagrangian formalism, but it did not lead to the appropriate evanescent 
behaviour of the eigenfunctions in the ICZ; instead they showed a large 
wavelength
oscillation. This comes from the fact that no control of the \bv frequency is 
possible in a Lagrangian formalism. Because of numerical truncation errors, 
the code
does not know if $N^2$ is slightly positive or negative in the ICZ. With an 
Eulerian formalism instead, $N^2$ appears explicitly in the movement 
equation,
and the correct value can be attributed to it. As was already pointed out by 
Dziembowski years ago (private communication), 
we emphasize thus that it is much better to use an Eulerian formalism for the 
finite difference scheme inside a deep
convective zone.

Our numerical method give us therefore the frequencies of the reflected and 
potentially excited modes by solving numerically, with the fully adiabatic 
code, the 'envelope problem' first. We then compute the damping rate of the 
modes taking into account the envelope and the core for which we made some 
valid approximations: the use of the quasi-adiabatic approximation and the 
eigenfunctions of the asymptotic theory.

\section[]{Results and discussion}

New samples of B supergiant stars have been observed and it is now commonly admitted that they present non-radial pulsations \citep{Waelkens1998,Aerts1999,Mathias2001,Saio2006,Lefever2007}. The Canadian MOST satellite observed the B supergiant HD\,163899 in June 2005 during 37 days and detected 48 frequencies $\lesssim$ 2.8 c/d with amplitudes of a few milli-magnitudes. Blue supergiants are however characterized by a radiative core in which the \bv frequency takes huge values. A strong radiative damping ensues and no such modes should be observed. \citet{Saio2006} have suggested that the presence of an ICZ around the core prevents some modes from entering the core. In that case the radiative damping is largely reduced and the $\kappa$-mechanism in the superficial layers can excite the mode. Some physical processes could however prevent that ICZ from developing; this is the case with mass loss and overshooting. We shall successively envision six different physical situations.

\subsection{A supergiant star without mass loss and without overshooting}

\begin{figure}
\begin{center}
\includegraphics[width=58mm,angle=270]{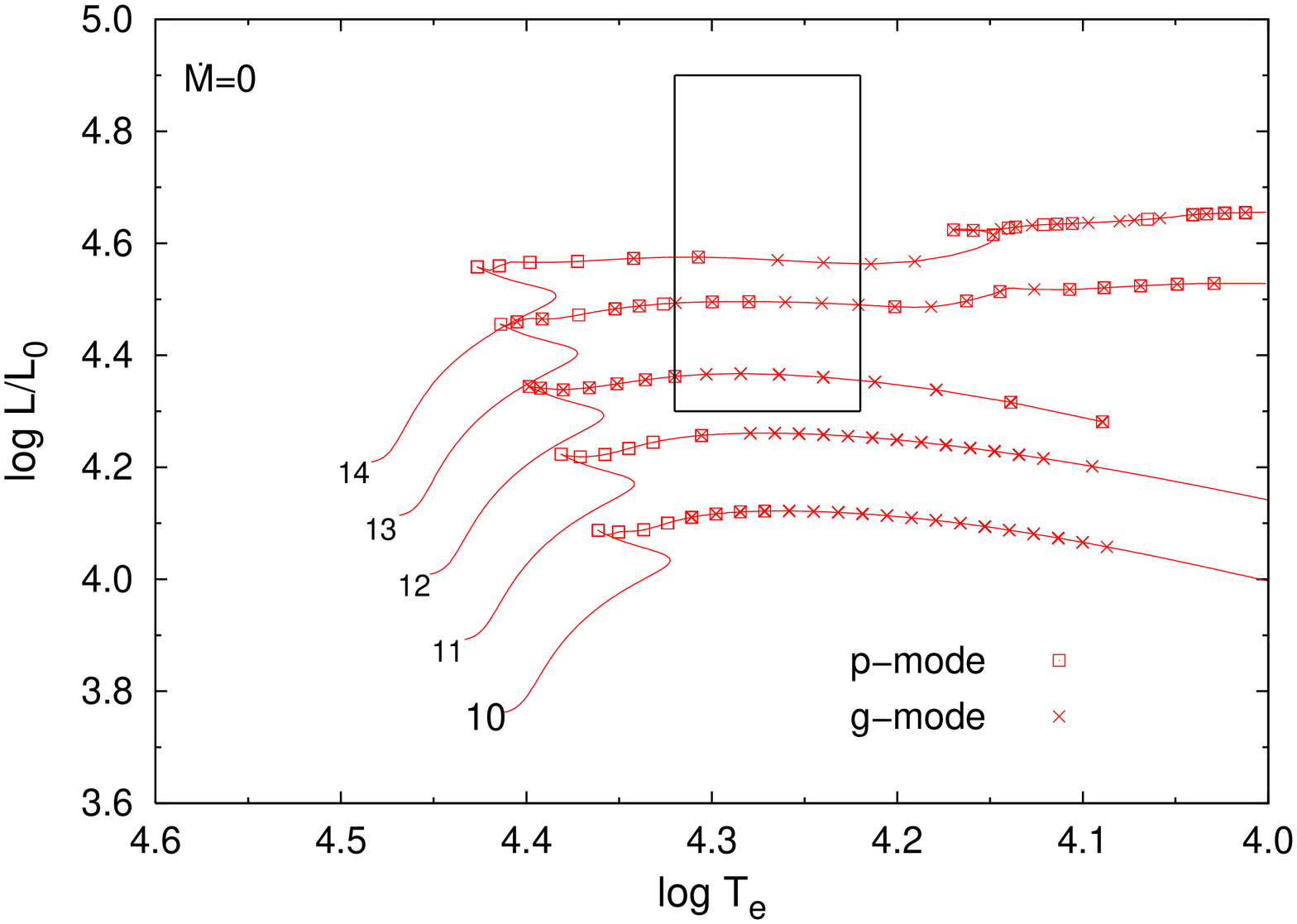}
\includegraphics[width=58mm,angle=270]{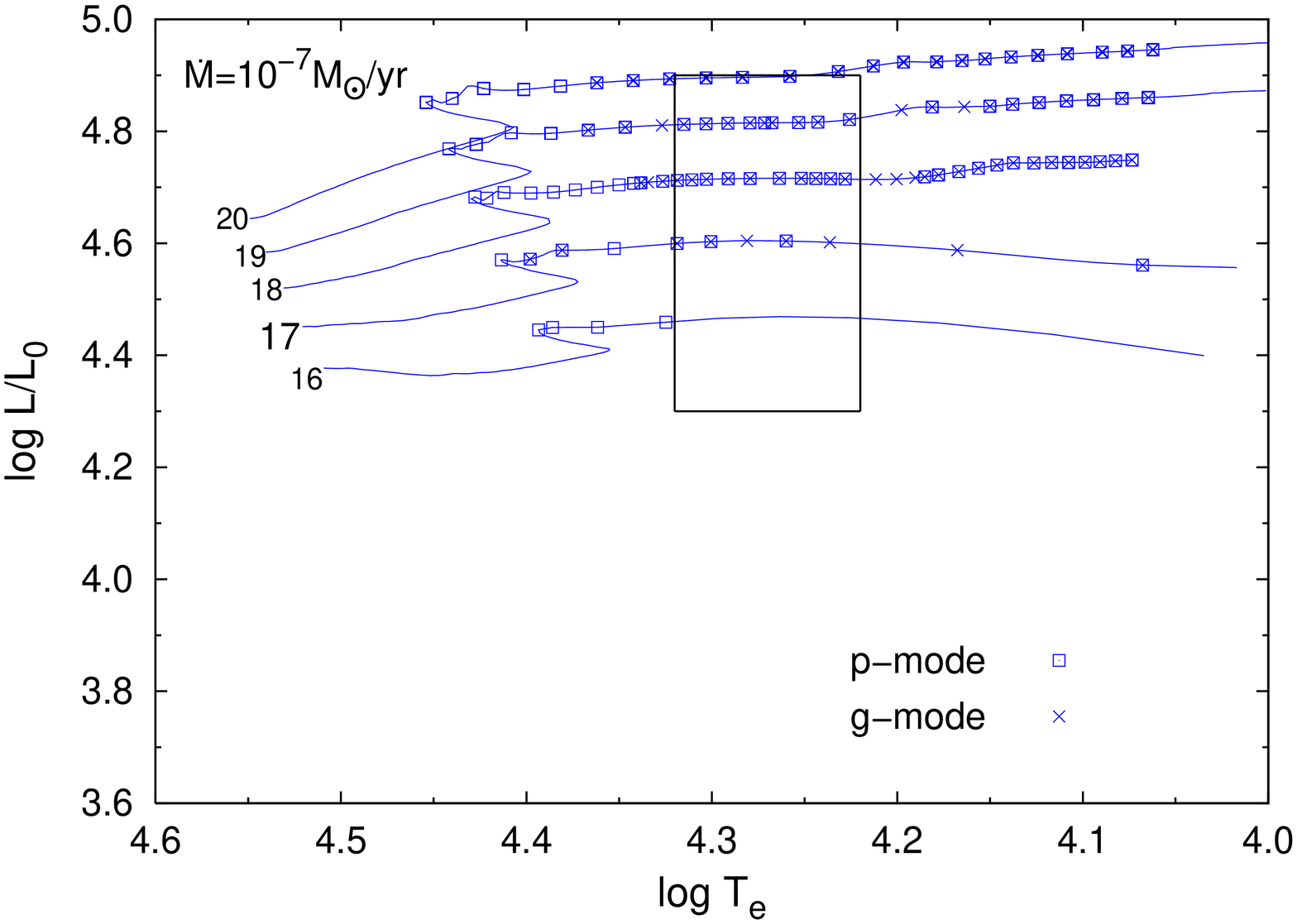}
 \caption{Evolutionary tracks and stability data in the HR diagram. The black box is the error box of the MOST star from \citet{Saio2006}. \textbf{Top} Without mass loss (from $10$ to $14\msun$). All sequences present excited g-modes (crosses) on the supergiant phase. \textbf{Bottom} With $\dot{M} = 10^{-7}\msun$/yr (from $16$ to $20\msun$): excited g-modes are present only for a mass higher or equal to $17\msun$.}
 \label{fig:hrbis}
\end{center}
\end{figure}

As mentioned, the excitation of the g-modes in supergiant stars is due to the opacity bump of iron in the surface layers thanks to the blocking effect of an ICZ (Sect.~\ref{sect:ICZ}). \citet{Saio2006} computed evolutionary models for a mass range of $7\leqslant M/M_{\odot}\leqslant20$ with an initial composition of (X,Z)=(0.7,0.02) and the OPAL opacities. Their models fit quite well the observations, though a frequency gap between p and g-modes still appears in the theoretical excited frequencies compared to the observed ones. They derived the estimated effective temperature from photometric analysis \citep{Schmidt1996}: for a B2Ib/II type star, $\log T_{\sf eff}$ is roughly between 4.22 and 4.32. The model which best reproduces the observed frequency range is a model of $15\msun$ at $\log T_{\sf eff}\approx{4.36}$ for which they assume that the frequency gap would be filled at least partially by rotationally splitted modes. We have computed models with mass from 10 to 18$\msun$. Fig.~\ref{fig:hrbis} shows evolutionary tracks with instability region for p and g-modes from $10$ to $14\msun$ (top panel). In agreement with the work of \citeauthor{Saio2006} we find excited g-modes on the supergiant phase.

\subsection{A supergiant star with mass loss and without overshooting}

The occurrence of stellar winds is ubiquitous for stars of almost all mass and constitutes
a key aspect of stellar physics. In terms of the mass loss integrated over the whole star life time,
it is most extreme in massive luminous stars, where the winds are almost certainly radiatively
driven. Most of hot massive stars (O, B, WR) indeed present a wind essentially due to the radiation
pressure: winds are driven by the resonant scattering of the stellar radiation \citep{Lamers1999}. 
When taking mass loss into account, the central temperature and the radiation pressure increase less rapidly in the course of central H-burning. As a result, the adiabatic temperature gradient decreases less and less as the mass loss rate increases. If a large enough mass loss rate is assumed during the MS, no ICZ is formed during the supergiant phase (Chiosi \& Maeder 1986). 

We adopted a mass loss rate of $10^{-7}\,\msun$/yr which is slightly larger than most observed MS mass loss rates in order to emphasize the effect of mass loss on the excitation of g-modes. Indeed, the mass loss recipe of \citet{Vink2000,Vink2001} gives mass loss rates of the order of $10^{-9}\,\msun$/yr to $10^{-8}\,\msun$/yr during the MS and of about $10^{-7}\,\msun$/yr on the post-MS phase for $14\msun$ and $16\msun$ stars. But the real mass loss rates are still under discussion until now (see for example \citealt{Bouret2005,Martins2005,Puls2006,Oskinova2007,Mokiem2007} and references therein).

\begin{figure}
\begin{center}
\includegraphics[width=65mm,angle=270]{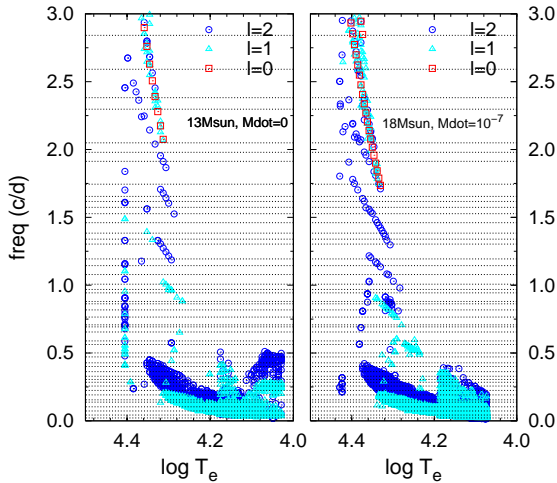}
 \caption{Frequency distribution of excited p and g-modes for supergiant models of $13\msun$ computed without mass loss (left panel) and of $18\msun$ computed with $\dot{M}=10^{-7}\msun$/yr (right panel). The horizontal dotted lines stand for the observed frequency of MOST \citep{Saio2006}.}
 \label{fig:Saio}
\end{center}
\end{figure}

We have computed evolutionary tracks with and without mass loss. Their location in the HR diagram is shown on Fig.~\ref{fig:hrbis}: tracks on top panel (resp. bottom panel) are computed without (resp. with) mass loss. We performed non-adiabatic computations to determine whether unstable modes were present or not. For the sequences computed without mass loss, there are indeed excited g-modes during the supergiant phase for models within the error box for the MOST star HD\,163899. Even at $10\msun$ the supergiant phase is characterized by excited g-modes. This is true for all higher masses. On the other hand, in the sequences computed with mass loss, we find excited g-modes only for stars initially more massive than about $17\msun$, i.e. with an instantaneous mass of $13.7\msun$ on the supergiant phase. With an even higher mass loss rate, this value becomes larger and larger.  These sequences without any excited g-modes differ by the absence of ICZ on the supergiant phase, due to the mass loss effect during the MS phase. Hence, all the modes enter the radiative core and suffer the strong radiative damping.
The frequency distribution of the theoretical excited modes is shown in the same kind of figure as in \citet{Saio2006} but here comparing without and with mass loss (Fig.~\ref{fig:Saio}). In both cases the frequency distribution of the theoretical excited modes is in good agreement with the observed frequencies.

\subsection{A supergiant star without mass loss and with overshooting}

\begin{figure}
\begin{center}
\includegraphics[width=65mm,angle=270]{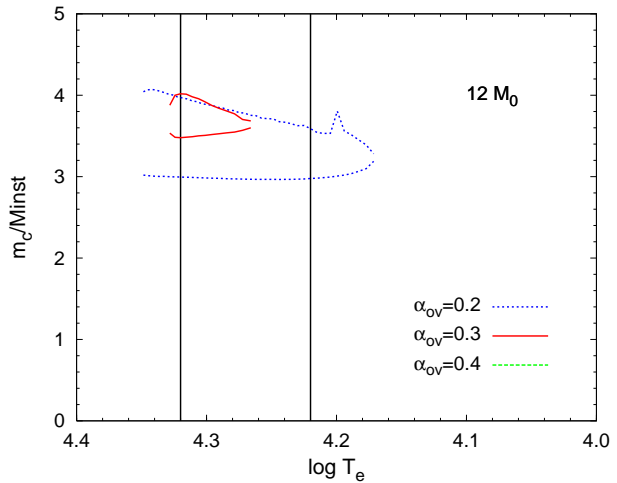}
 \caption{Evolution of the mass extension of the ICZ during the supergiant phase. On the x-axis we have the effective temperature decreasing after the MS. Each point of this axis is a model. For each model the mass fraction of the limits of the convective zone are indicated on the y-axis.
For a small amount of overshooting $\alpha_{\sf ov}=0.2$, the ICZ is well develops, it is much smaller for $0.3$ and it disappears for $0.4$.}
 \label{fig:ZC2}
\end{center}
\end{figure}

The convective stability criterion fixes the boundary of the convective core as the layer where the buoyancy acceleration vanishes. The velocity of the globule is however different from zero and the globule can penetrate the radiative region. Such an enlargement of the mixed core makes the star more luminous, it also increases the core H burning lifetime while, in the HR diagram, the MS track is more extended. 
We model here overshooting by just extending the mixed region but keeping $\nabla=\nabla_{\sf rad}$ in the overshooting region. When assuming a large enough overshooting during the MS, the MS 'neutral' region (where $\nabla_{\sf rad}\cong\nabla_{\sf ad}$) does not exist above the convective core and this can prevent the formation of an ICZ during the post-MS phase. No g-modes are thus found to be excited. We have checked the presence of an ICZ by progressively increasing the overshooting parameter from one evolutionary sequence to the next (Fig.~\ref{fig:ZC2}). For a model of $12\msun$ the ICZ is well develop for a small amount of overshooting $\alpha_{\sf ov}=0.2$, it is much smaller for $0.3$ and it completely disappears for $0.4$. In the cases considered here, the overshooting is however small enough to keep the error box well within the post-MS phase.

\subsection{A main sequence star without mass loss and with overshooting}

\begin{figure}
\begin{center}
\includegraphics[width=58mm,angle=270]{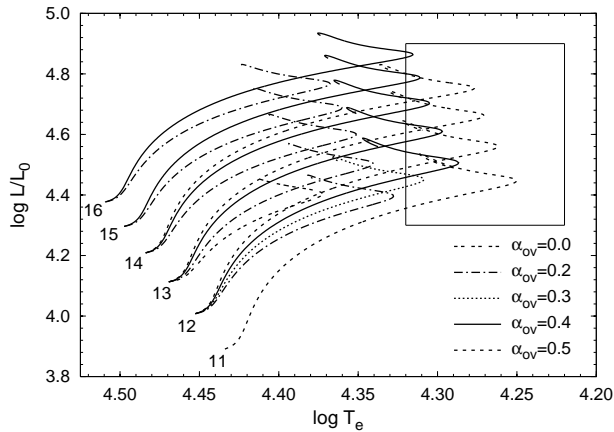}
\caption{Main sequence evolutionary tracks computed with different overshooting parameters ranging from $\alpha_{\sf ov}=0.2$ to $0.5$. The black box is the error box of HD 163899. Main sequence evolutionary tracks with at least $\alpha_{\sf ov}=0.3$ cross this error box.}
 \label{fig:hrov}
\end{center}
\end{figure}

Another possibility to solve the problem of the presence of excited g-modes would be to bring MS evolutionary tracks into the error box of HD\,163899. This could be achieved by including larger overshooting in MS models. Indeed, with overshooting, the MS phase of evolution reaches lower effective temperatures. We computed evolutionary tracks with different overshooting parameter, ranging from $\alpha_{\sf ov}=0.2$ to $0.5$ (Fig.~\ref{fig:hrov}). MS evolutionary tracks cross the error box for an overshooting parameter equal to or larger than $0.3$. 

\begin{figure}
\begin{center}
\includegraphics[width=65mm,angle=270]{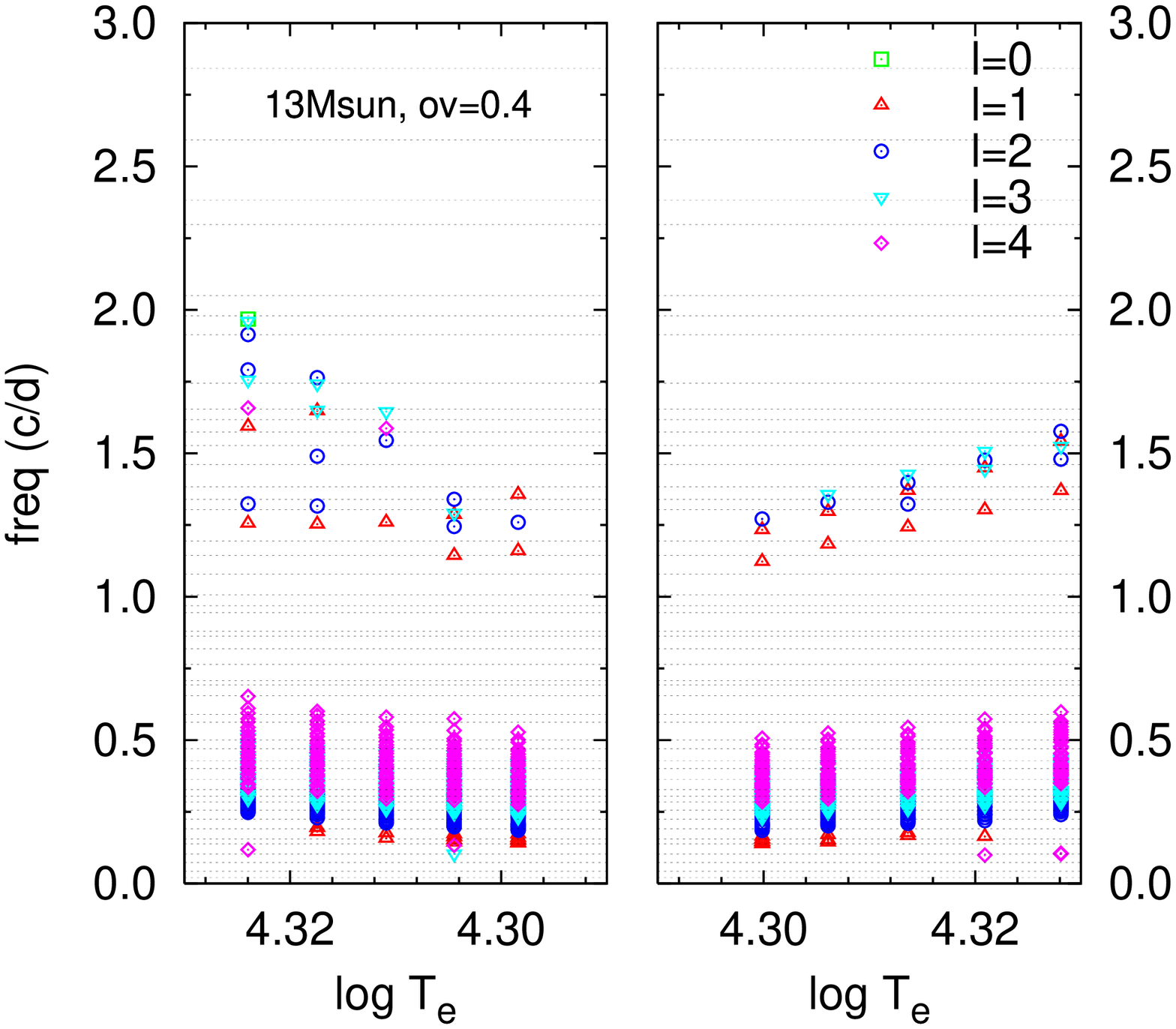}
\includegraphics[width=65mm,angle=270]{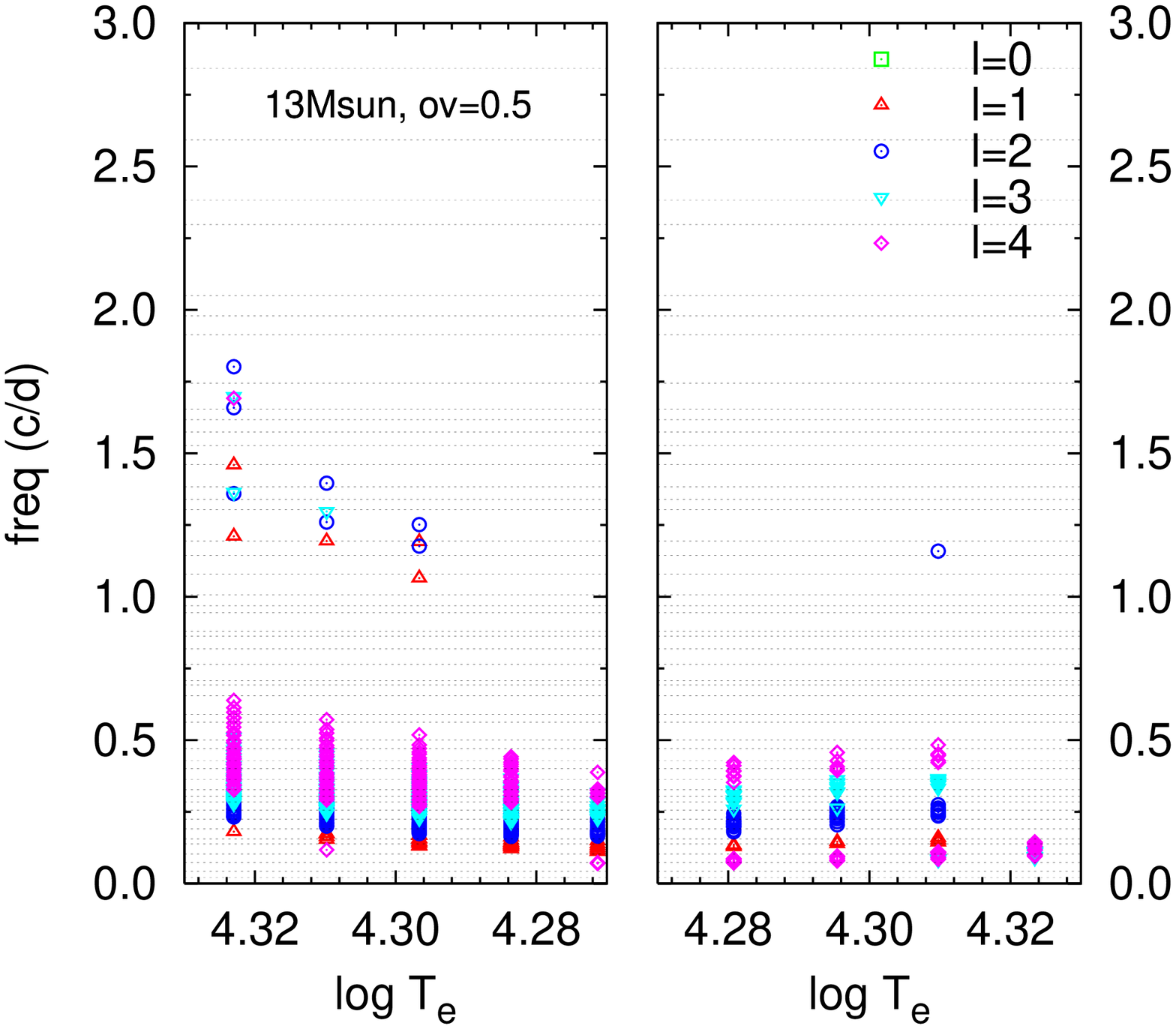}
\caption{Frequency distribution of the excited p and g-modes during the MS (decreasing $T_{\sf eff})$ and near the turn off (increasing $T_{\sf eff}$) for $13\msun$ models computed with overshooting. The horizontal dotted lines stand for the observed frequency of MOST \citep{Saio2006}. The agreement between the observed frequencies and the theoretical mode spectrum is not as good as it was for the supergiant models. \textbf{Top} $\alpha_{\sf ov}=0.4$. \textbf{Bottom} $\alpha_{\sf ov}=0.5$.}
 \label{fig:freq13-ov.4-5}
\end{center}
\end{figure}

On the MS phase, massive stars present a convective core surrounded by a radiative envelope. The \bv frequency is therefore zero in the core and no damping can occur. The $\kappa$-mechanism due to the Fe-opacity bump at $\log T=5.2$ excites p and g-modes.
We performed non-adiabatic computations which revealed excited g-modes in all the sequences. The spectrum of the theoretical excited modes is shown on Fig.~\ref{fig:freq13-ov.4-5} for $\alpha_{\sf ov}=0.4$ and 0.5 during the MS (decreasing effective temperatures) and near the turn off (increasing effective temperatures). The agreement in the mode spectrum is however not as good as it was for the supergiant model.

\subsection{A supergiant star in the helium burning phase}

Blue supergiants could also be He-burning stars. Their structure shows a small convective core surrounded by a radiative envelope in which an ICZ may still exist. As long as such an ICZ is present g-modes can still be excited.
After the disappearance of the ICZ, the radiative damping is no longer reduced by this reflective barrier and g-modes are all stable. In a forthcoming paper we shall analyze the excitation of g-modes in He burning models.

\subsection{Ledoux's criterion}

\begin{figure}
\begin{center}
\includegraphics[width=65mm,angle=270]{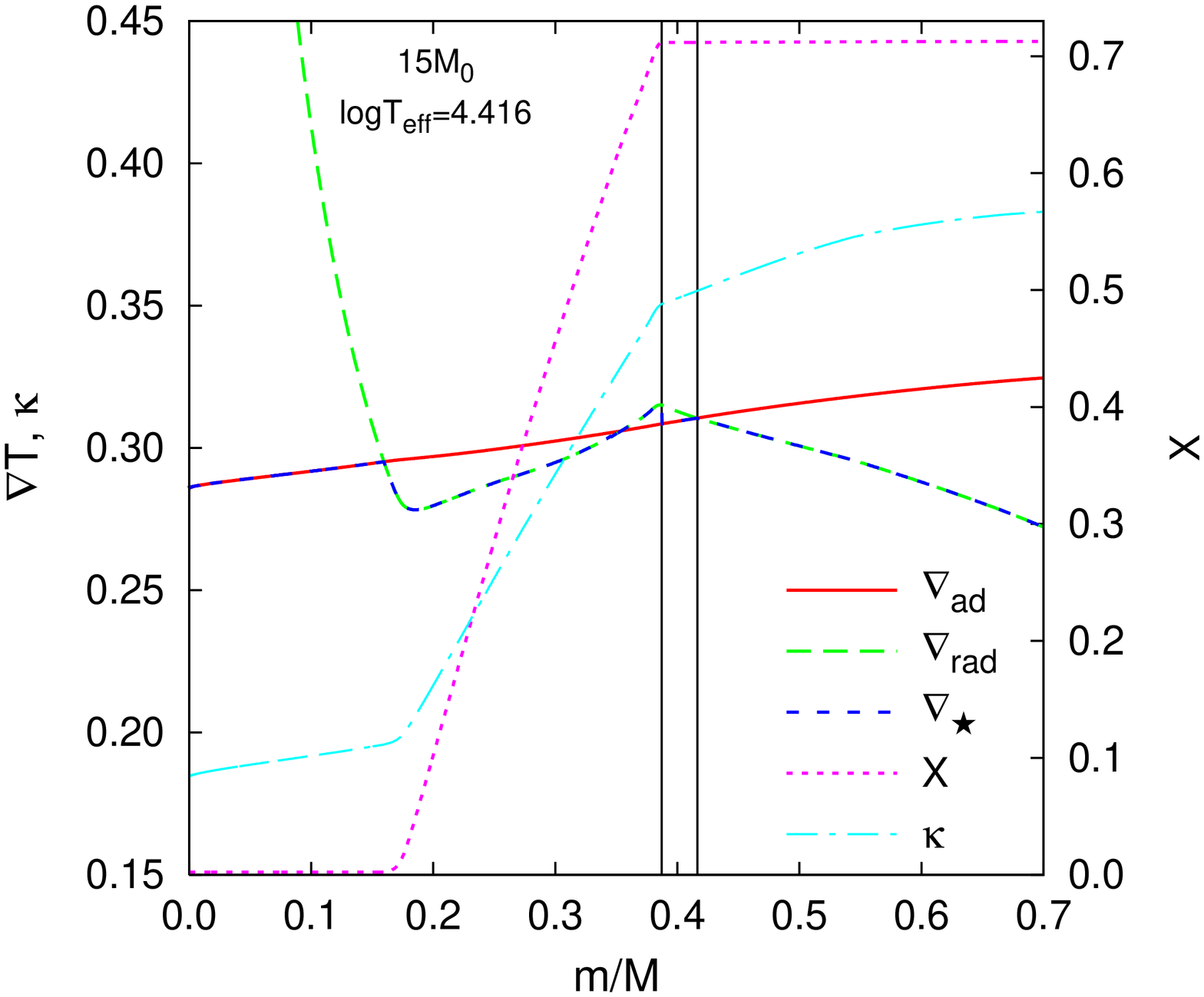}
\includegraphics[width=65mm,angle=270]{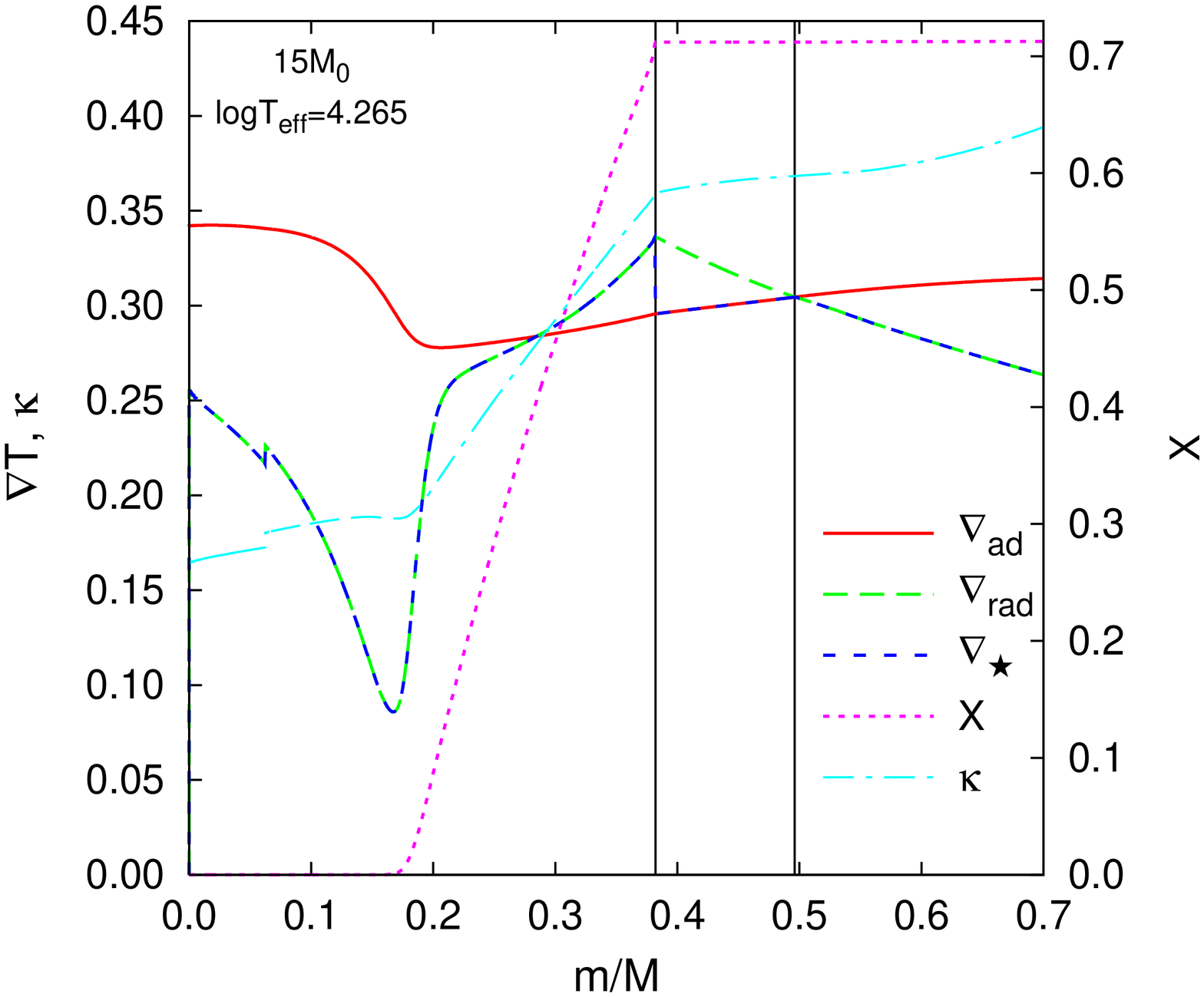}
 \caption{Temperature gradients, opacity, $\kappa$, and hydrogen abundance, X, in a model of $15\msun$ computed with the Ledoux's criterion for convective instability. The vertical solid lines stand for the boundary of the ICZ. \textbf{Top} MS model. \textbf{Bottom} Post-MS model.}
 \label{fig:Ldx}
\end{center}
\end{figure}

The hydrogen rich layers surrounding the core during the MS phase are potentially unstable towards convection and a partial or semi convective mixing is supposed to take place in order to achieve neutrality versus the convective criterion. This criterion is either the Schwarzschild criterion or the Ledoux criterion and we choose here the Schwarzschild one for which a region is convective if $\nabla_{\sf rad}\geq \nabla_{\sf ad}$. 
However, adopting the Ledoux's criterion ($\nabla_{\sf rad} = \nabla_{\sf L} = \nabla_{\sf ad}+ (\beta/(4-3\beta))\nabla_{\sf \mu}$, where $\nabla_{\sf \mu}={\rm d}\ln{\mu}/{\rm d}\ln P$ and $\beta$ is the gas pressure to total pressure ratio) alters the structure of the star since it adds to the adiabatic temperature gradient the mean molecular weight gradient which is formed on top of the convective core during MS. The MS convective core is therefore not altered by the choice of criterion ($\nabla_{\sf \mu}=0$), but adopting Ledoux's criterion affects the size and the location of the ICZ on the post-MS \citep{Lebreton}. During the MS, the presence of the mean molecular weight gradient prevents the formation of a region in which the neutrality of gradients is reached ($\nabla_{\sf rad}\cong \nabla_{\sf ad}$). However, outside the core, the radiative gradient increases due to a larger opacity resulting from the increase in the hydrogen abundance.  As a result, an ICZ appears above the region of variable mean molecular weight (Fig.~\ref{fig:Ldx}, top panel). This ICZ develops in an homogeneous region where no nuclear reactions take place. The bottom of the ICZ corresponds to the base of the constant hydrogen rich envelope, which corresponds to the maximum extent of the MS convective core. On the post-MS phase this ICZ remains at the same location, well outside the nuclear burning shell (Fig.~\ref{fig:Ldx}, bottom panel).

\section{Conclusions}

The formation of an ICZ during the post-MS phase is a result of the MS evolution. The presence of excited g-modes in B supergiant stars depends therefore on physical processes during the MS: overshooting, mass loss, convective criterion. We show that with significant mass loss or overshooting during the MS no ICZ appears during the post-MS phase. In a future work we shall extend this preliminary analysis in order to define an instability strip depending on those physical aspects and compare it to the observations. In this confrontation process, it is clear that asteroseismology of massive supergiant stars can give us a better understanding on the physical processes not only during the supergiant phase but also during the MS phase.

\label{lastpage}

\end{document}